\documentclass{elsart}

\usepackage{natbib}

 \usepackage{epsfig}

\usepackage{amssymb}

\begin{document}

\begin{frontmatter}

\title{Dark Matter Searches with GLAST}

\author{E.Nuss\corauthref{cor}
}
\address{LPTA Montpellier II University}
\corauth[cor]{Corresponding author}
\ead{eric.nuss@lpta.in2p3.fr}

\author{on behalf of GLAST LAT Dark Matter and New Physics WG}

\begin{abstract}

The Gamma-Ray Large Area Space Telescope (GLAST), 
scheduled to be launched in fall 2007, is the next generation satellite 
for high-energy gamma-ray astronomy. The Large Area Telescope (LAT),
 GLAST main instrument, with a wide field of view ($>$ 2 sr), 
a large effective area  ($>$ 8000 cm$^2$ at 1 GeV) and 20 MeV - 300 GeV energy range, 
will provide excellent 
high energy gamma-ray observations for Dark Matter searches.
In this paper we examine the potential of the LAT to detect gamma-rays 
coming from WIMPS annihilation in the context of supersymmetry.
As an example, two search regions are investigated: 
the galactic center and the galactic satellites.

\end{abstract}

\begin{keyword}
Dark Matter \sep New Physics \sep Gamma-rays

\end{keyword}

\end{frontmatter}

\section{Introduction}

The Cold Dark Matter (CDM) problem is one of the most fascinating
 and intriguing issue in present day cosmology.
It has been a subject of special interest to 
high-energy physicists, astrophysicists and cosmologists for many years. 
According to a wealth of observations and arguments,
such as excessive peculiar velocities of galaxies within
 clusters of galaxies or gravitational arcs, 
 it can make up a significant fraction of the mass of the universe. 
On the galactic scale, dark matter halos are required to
 explain the observed rotation curves in spiral galaxies
or the velocity dispersion in elliptical galaxies.
Virtually all proposed candidates require physics beyond the standard model
 of particle physics and could be detected through stable products 
of their annihilations:  
energetic neutrinos, antiprotons, positrons, gamma-rays etc.
Supersymmetric extensions of the standard model of
 particle physics provide a natural candidate for CDM in the form 
of a stable uncharged Majorana fermion (neutralino).

Hereafter, we briefly report on the potential of the GLAST 
high-energy gamma-ray telescope
to detect Dark Matter  indirectly through their annihilation 
in the halo of the galaxy and  in
clumpy substructures throughout the Galactic halo.

\section{The Gamma-ray Large Area Space Telescope (GLAST) mission}

GLAST is the next-generation gamma-ray telescope 
for studying high-energy gamma-ray emission
from astrophysical sources
\footnote{For more details, see the GLAST website at: http://glast.gsfc.nasa.gov/}.
Its main instrument, the Large Area Telescope (LAT), is
a modular 4x4-tower pair-conversion telescope  
instrumented with a plastic anticoincidence shield which vetoes charged cosmic rays,
a tracker of silicon strip planes with foils of tungsten converter
followed by a segmented CsI electromagnetic calorimeter.
A photon traversing the tracker will have high probability of converting into the 
tungsten foils, thus
forming an electron-positron pair, subsequently tracked by the silicon strip detectors.
 The reconstructed
trajectories of this pair, together with their energy deposition in the calorimeter,
 allows to reconstruct the 
direction and energy of the incident gamma-ray photon.
The main characteristics of the LAT detector, i.e. the effective area,
 point spread function and energy dispersion, have been obtained from detailed
 Monte Carlo studies and parameterized by a series of functions: 
the Instrument Response Function (IRF)
\footnote{http://www-glast.slac.stanford.edu/software/IS/glast\_lat\_performance.htm}.
The main LAT Science Performances relevant for Dark Matter searches are summarized in
Table \ref{table1}.
\begin{table}
\caption{Summary of the main LAT Science Performances relevant for Dark Matter searches}
\begin{center}
\begin{tabular}{|ll|}
\hline
Parameter                               
& Current Best Estimate \\
\hline
Peak Effective Area (in range 1-10 GeV) 
& $\simeq\ 9000$ cm$^2$\\
Energy Resolution 10 GeV on-axis        
& $<$ 6 \% \\
PSF 68\% 10 GeV on-axis                 
& $<$ 0.1$^0$ \\
Field of View                           
& $>$ 2 Sr\\
Dead Time                               
& 26.5 $\mu$sec/event nominal\\
\hline
\end{tabular}
\end{center}
\label{table1}
\end{table}
The LAT takes much of its basic design concept from its predecessor EGRET but the
energy range (20 MeV-300 GeV and above), field-of-view (greater 
than 2 steradians) and angular resolution will provide the LAT with a
 factor $\sim$30 better sensitivity.
This improvement should allow to detect several thousands of new
 high-energy sources and shed light on many issues left open by EGRET.

A detailed description of GLAST science prospects and an introduction to
 the experiment can be found in \cite{GLAST_PM}. 
The LAT is now completed and handed to the
spacecraft vendor for integration.
It represents the largest silicon strip 
detector ever built for space applications. 

GLAST is scheduled for launch in fall 2007.

\section{Dark Matter searches with GLAST at the Galactic Center}

\begin{figure}
\begin{center}
\includegraphics*[width=14cm,angle=0]{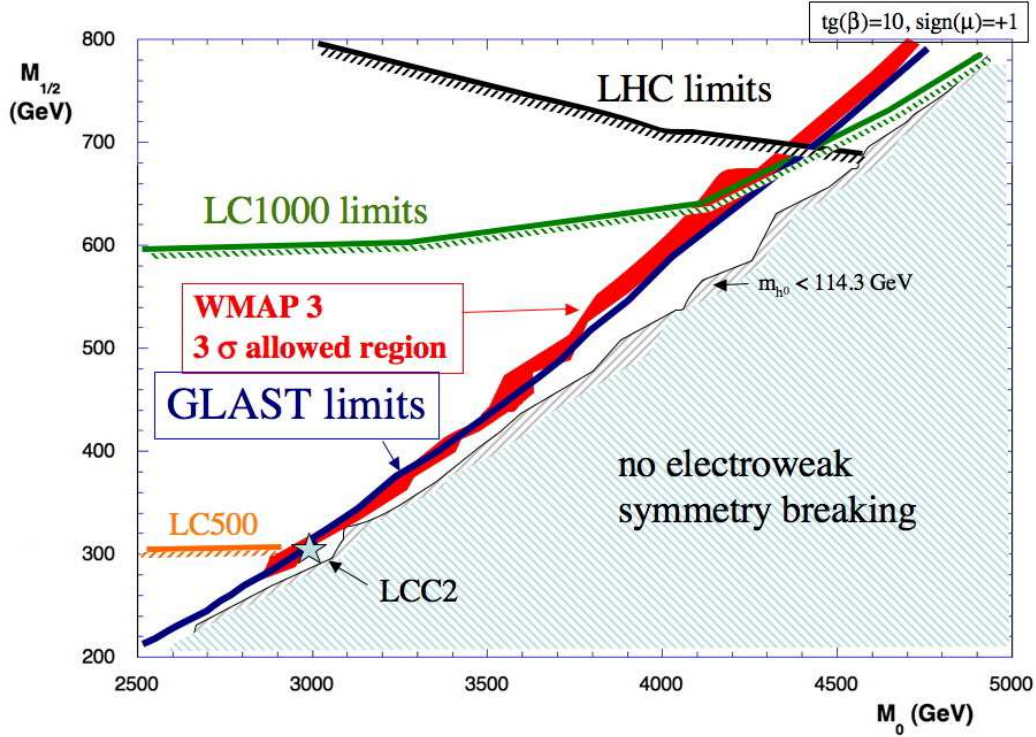}
\label{tg10}
\end{center}
\caption{
GLAST $5\ \sigma$ sensitivity to Dark Matter signal for 5 years observation of 
the Glactic Center in the
mSUGRA ($m_0$,\ $m_\frac{1}{2}$ ) plane
for $tan(\beta)=10$, $A_0=0$, $\mu>0$ and top quark mass $m_{t}=174$ GeV.
}
\end{figure}

\begin{figure}
\label{tg55}
\begin{center}
\includegraphics*[width=14cm,angle=0]{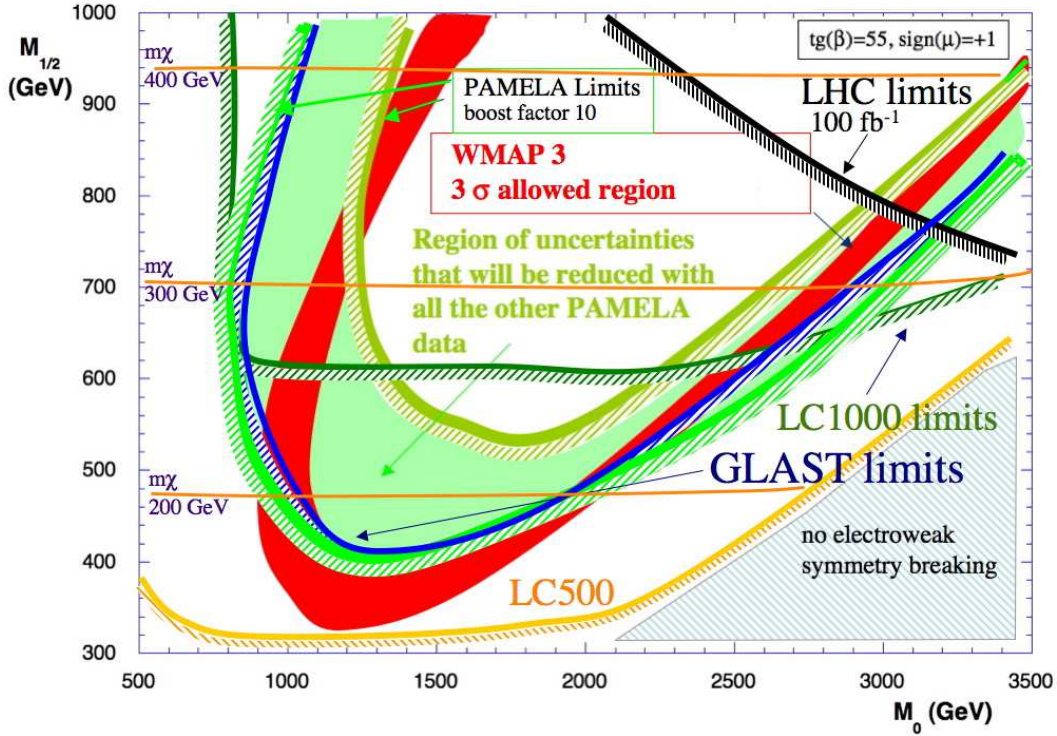}
\end{center}
\caption{
GLAST $5\ \sigma$ sensitivity to Dark Matter signal for 5 years observation of 
the Glactic Center in the
mSUGRA ($m_0$,\ $m_\frac{1}{2}$ ) plane
for $tan(\beta)=55$, $A_0=0$, $\mu>0$ and top quark mass $m_{t}=174$ GeV.
}
\end{figure}

The unusual point source found by EGRET at or near the Galactic Center
gives a strong indication of an excess with
respect of the standard model production of gamma-rays. 
The spectrum of this potential source of WIMP annihilation
is extremely hard, non-power law, and the source
is possibly extended (Mayer-Hasselwander, et.al. 1998). 
Several hypotheses for the origin of this source have been proposed
and the dark matter hypothesis is not yet ruled out. 

The figures (1) and (2) shows 
the mSUGRA ($m_0$,\ $m_\frac{1}{2}$ ) plane
with parameters similar to the two mSUGRA benchmark points LCC2 and LCC4 
respectively, as defined in \cite{Baltz}.
The GLAST $5\ \sigma$ sensitivity to dark matter signal (continuum spectra)
for 5 years observation of a $0.1^0$
region around the galactic center is 
shown at the blue lines and below. 
The diffuse background was estimated using the 
GALPROP code v50
\footnote{http://galprop.stanford.edu/na\_home.html}
(point source substracted).
The GLAST sensitivity is compared with 
mSUGRA accelerator limits from \cite{accel}. 
The dark matter halo used to compute the GLAST indirect search sensitivity is a 
truncated Navarro Frank and White (NFW) halo profile (as defined in \cite{Baltz}).
The methodology for deriving the various curves is presented in \cite{Cesarini}.
The PAMELA antiprotons limits are from \cite{Lionetto}.
In these figures, it can be seen that GLAST can explore
a good portion of the supersymmetric parameter space for this kind of halo.

\section{Dark Matter searches with GLAST in Galactic Satellites}

According to N-body simulations on test particles with only gravitational interactions 
(see \cite{clumpsNavarro}, \cite{clumpsMoore}),
WIMPs are expected to form high density dark matter substructures in the galactic halo
with masses less than approximately $10^{7-8}$ solar masses.

The number of Milky Way dark matter satellites 
observable by GLAST have been estimated 
using a semianalytic method \cite{TaylorAndBabul} where
the dark matter satellite distribution is roughly spherically
symmetric about the galactic center and extends well beyond the 
solar orbit; thus the dark matter satellites are located mostly at
 high galactic latitudes. 

The figure 
(3) 
shows 
the number of dark matter clumps observable by GLAST for 5 years observation for 
two benchmark mSUGRA points as a function of the significance.
The black line corresponds to simulations for
the LCC4 mSUGRA model defined in \cite{Baltz} 
and the grey line corresponds to the LCC2 model.
The LCC1 and LCC3 models are much less favorable, with no clumps detected.
The background was estimated using the EGRET point 
source subtracted sky map above 1 GeV.

\begin{figure}
\label{nclumps}
\begin{center}
\includegraphics*[width=14cm,angle=0]{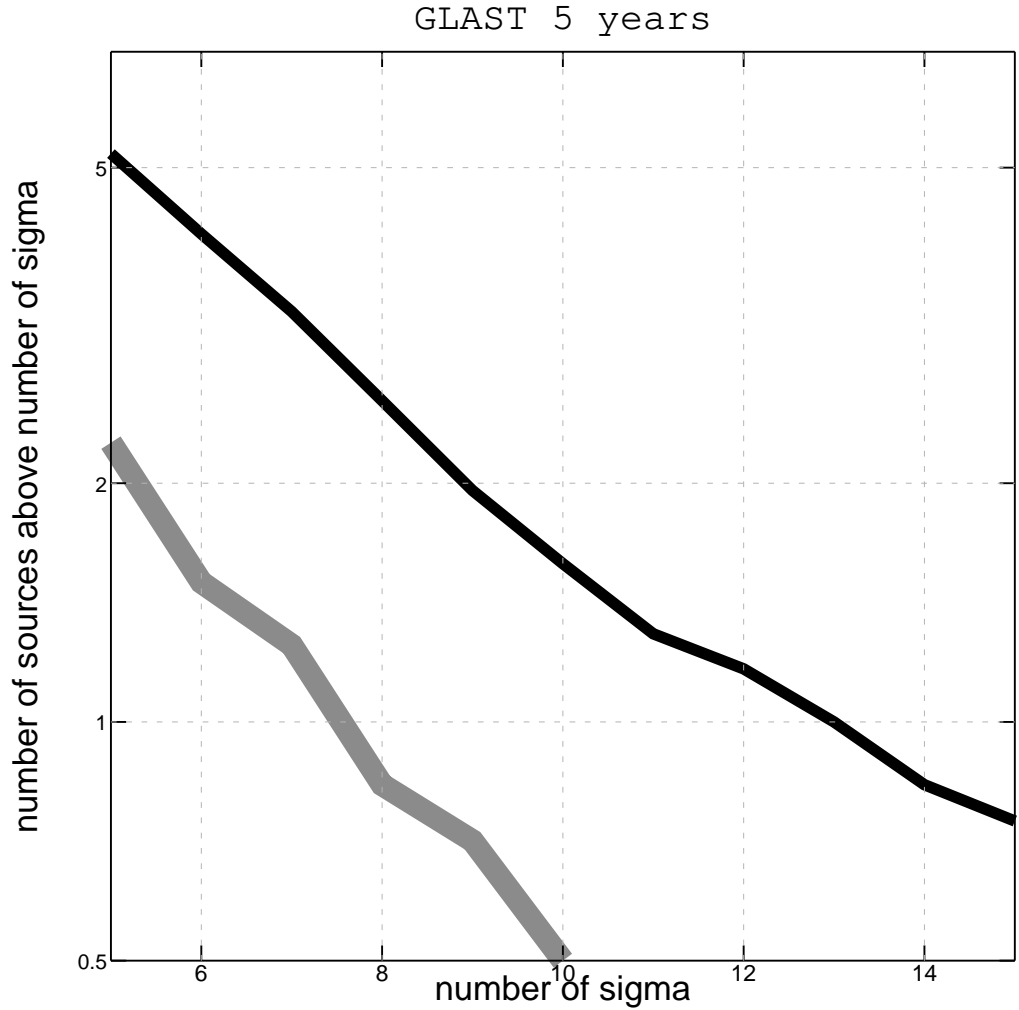}
\end{center}
\caption{Number of clumps observed by GLAST vs. number of sigma significance in 5 years of all-sky scanning.}
\end{figure}

\section{Conclusions}
\label{Section 3}

In this paper we presented a preliminary GLAST-LAT sensitivity to
indirect gamma-ray signature of WIMP annihilation,
based on the best simulations currently 
available to the LAT collaboration.
As an example, two search regions have been investigated, 
the Galactic Center and the galactic satellites.
Despite its dependence 
on a still unprecisely known background, 
our estimate shows that the GLAST telescope,
should be capable of searching for Dark Matter in the energy range 
$E_\gamma \ge 20$ MeV.
The peculiar timing, spectral, and spatial signatures
associated to WIMP annihilations will 
permit to reject other possible astrophysical sources of gamma-rays.


\begin{thebibliography}{}

\bibitem[Baer et al.(2004)]{accel} H. Baer, A. Belyaev, T. Krupovnickas et al. , JCAP 0408 (2004) 005
\bibitem[Baltz et al.(2006)]{Baltz} E. A. Baltz, M. Battaglia, M. E. Peskin et al, hep-ph/0602187 
\bibitem[Cesarini et al.(2004)]{Cesarini} A.Cesarini, F.Fucito, A.Lionetto et al,, Astroparticle Physics 21, 267-285, 2004.
\bibitem[Lionetto et al.(2005)]{Lionetto} A.Lionetto, A.Morselli, V.Zdravkovic et al., Astropart. Phys. JCAP09(2005)010.
\bibitem[Mayer-Hasselwander et al.(1998)]{Mayer} H. A. Mayer-Hasselwander et al., Astron. $\&$ Astrophys. 335, 161 (1998).
\bibitem[Michelson (2001)]{GLAST_PM}
P.F. Michelson, ``The Gamma-ray Large Area Space Telescope Mission: 
Science Opportunities'' in {\em AIP Conf. Proc. 587: Gamma 2001 : Gamma-Ray Astrophysics}, 2001, pp.713-+.
\bibitem[Moore et al.(1999)]{clumpsMoore} B. Moore, T. Quinn, F. Governato, J. Stadel and G. Lake, Mon. Not. R. Astron. Soc. 310, 1147 (1999).
\bibitem[Navarro et al.(1997)]{clumpsNavarro} J. F. Navarro, C. S. Frenk and S. D. M. White, Astrophys. J. 490, 493 (1997). 
\bibitem[Taylor and Babul(2004)]{TaylorAndBabul} J. E. Taylor and A. Babul, Mon.Not.Roy.Astron.Soc.364, 2005 and references therein

\end{thebibliography}
\end{document}